\documentclass[aps,pra,twocolumn,byrevtex,showpacs,floatfix,superscriptaddress,nofootinbib]{revtex4-1}%,preprint,draft
\pdfoutput=1
\usepackage{graphicx}
\usepackage{dcolumn}
\usepackage{amsmath}
\usepackage{epsfig}
\usepackage{bm}
\usepackage{amssymb}
\usepackage{hyperref}
\hypersetup{colorlinks=true,linkcolor=blue,citecolor=blue,urlcolor=blue}

  \newcommand{\ket}[1]{|#1\rangle}

  \DeclareMathOperator{\sinc}{sinc}

\begin{document}

\title{Proposal for Automated Transformations in Single-Photon Multipath Qudits}

\author{R. D. Baldij\~ao}
\affiliation{Departamento de F\'isica, Universidade Federal de Minas Gerais,Belo Horizonte, MG 31270-901, Minas Gerais, Brazil}
\affiliation{Instituto de F\'isica Gleb Wataghin, Universidade Estadual de Campinas, Campinas, SP 13083-859, Brazil}

\author{G. F. Borges}
\affiliation{Departamento de F\'isica, Universidade Federal de Minas Gerais,Belo Horizonte, MG 31270-901, Minas Gerais, Brazil}

\author{B. Marques}
\affiliation{Instituto de F\'isica, Universidade de S\~ao Paulo,  SP 05315-970, Brazil}

\author{M. A. Sol\'is-Prosser}
\affiliation{Center for Optics and Photonics and MSI-Nucleus on Advanced Optics,
Universidad de Concepci\'on, Casilla 4016, Concepci\'on, Chile}
\affiliation{Departamento de F\'isica, Universidad de Concepci\'on, Casilla 160-C, Concepci\'on, Chile}

\author{L. Neves}
\affiliation{Departamento de F\'isica, Universidade Federal de Minas Gerais,Belo Horizonte, MG 31270-901, Minas Gerais, Brazil}

\author{S. P\'adua}
\affiliation{Departamento de F\'isica, Universidade Federal de Minas Gerais,Belo Horizonte, MG 31270-901, Minas Gerais, Brazil}

\date{\today}

\begin{abstract}
We propose a method for implementing automated state transformations on single-photon multipath qudits encoded in a one-dimensional transverse spatial domain. It relies on transferring the encoding from this domain to the orthogonal one by applying a spatial phase modulation with diffraction gratings, merging all the initial propagation paths with a stable interferometric  network, and filtering out the unwanted diffraction orders. The automated feature is attained by utilizing a programmable phase-only spatial light modulator (SLM) where properly designed diffraction gratings displayed on its screen will implement the desired transformations, including, among others, projections, permutations and random operations. We discuss the losses in the process which is, in general, inherently nonunitary. Some examples of transformations are presented and, considering a realistic scenario, we analyse how they will be affected by the pixelated structure of the SLM screen. The method proposed here enables one to implement much more general transformations on multipath qudits than it is possible with an SLM alone operating in the diagonal basis of which-path states. Therefore, it will extend the range of applicability for this encoding in high-dimensional quantum information and computing protocols as well as fundamental studies in quantum theory.
\end{abstract}

\maketitle

\section{\label{sec:Int} Introduction}

The physically allowed transformations of quantum states, namely, quantum operations, are one of the basic requirements for any task in quantum information processing and computing \cite{NIELSEN}. They are also at the core of many fundamental issues in quantum theory, for instance, the formulation of contextuality in terms of state transformations \cite{Spekkens2005,Mazurek2016}. 

The optical approaches on these phenomena require the ability to implement operations on the photonic degree of freedom used to encode information.  If, on one hand, the single-photon polarization offers the simplicity for dealing with the state transformations, on the other hand, it is constrained to lie in a two-dimensional Hilbert space, thus ruling out any possibility of accessing higher-dimensional spaces with single-photon states. That is, with this encoding one is restricted to qubit-based applications only, losing the many advantages that can be exploited from high-dimensional encodings \cite{Fujiwara03, Cerf02, Vertesi10}. However, using the transverse spatial profile of a single-photon multimode field, there is, in principle, no limit to the information content it can carry out. This can be seen by decomposing the field profile into any infinite orthonormal discrete basis of functions, such as Hermite-Gauss or Laguerre-Gauss functions \cite{Walborn2010}, among others. In particular, the Laguerre-Gauss decomposition shows that one can encode information in the infinite-dimensional orbital angular momentum (OAM) states of single photons \cite{Krenn2016}. Restricting this encoding to finite $D$-dimensional subspaces, one generates OAM qudit states which is, currently, one of the main experimental approaches to investigate practical and fundamental issues in quantum mechanics for high-dimensional systems \cite{BoydOAM1,Berkhout2010,Gonzalez06,zeilingerOAM,AWhiteOAM}.

Another successful and, perhaps, simpler approach to encode information in high-dimensional spaces using the photonic spatial degree of freedom is to split its transverse profile into a finite set of $D$ distinguishable spatial modes of propagation. This can be done in many ways, for instance, with a multiport beam-splitter \cite{UnitaryZeilinger}, an array of slits (or pinholes) \cite{LEOFENDAS,SteveQKD}, a multicore optical fiber \cite{OSulivanHale05,YunhongQKDFiber,CanasQKDFiber}, and so on. This type of encoding, which we shall refer as a \emph{multipath qudit}, may be implemented either in a one-dimensional (1D) spatial domain (e.g., with an array of slits) or in a two-dimensional spatial domain (e.g., with a multicore fiber). Here, we will consider only 1D single-photon multipath qudits. 

Recently, the use of programmable spatial light modulators (SLM) has enhanced the potential for this encoding, providing advances ranging from automated state preparation \cite{LEOMIGUEL} to automated state transformations \cite{Lima:09}. In turn, these advances provided a fertile ground for many recent applications of these multipath qudits such as the demonstration of novel quantum tomographic techniques \cite{Lima:11,WAWA}, quantum algorithms \cite{Breno12}, entanglement characterization \cite{ALEJA} and concentration \cite{Breno2013}, simulation of decoherence \cite{BRENOSLM},  quantum key distribution \cite{Etcheverry2013}, state discrimination \cite{MiguelMario}, and contextuality tests \cite{Canas2016,Arias2015,Canas2014}. However, regarding the transformations (in which the final state is preserved), so far the operations implemented via SLMs have been restricted to diagonal ones in the basis of which-path states. In addition, they are, in general, state dependent, i.e., if one wants to implement a given transformation on the qudit, the SLM must be configured differently depending on the input state \cite{Lima:09,Breno2013}.  These features, altogether, are limiting and unwanted if more general transformations are required: for instance, transformations  with non-zero off-diagonal elements could be used to simulate quantum jumps \cite{BRENOSLM}; state-independent operations preserving the final state would be suitable to implement sequential transformations  \cite{ADANELIAS,BRENOHardy}. 

With the goal of extending even more the range of applicability for these 1D single-photon multipath qudits, in this work we propose a method for implementing automated transformations on their states, which will be much more general than the current ones described above. Our method relies on transferring the encoding from the 1D spatial domain, say $x$, to the orthogonal one, $y$. This will be accomplished by applying a spatial phase modulation with diffraction gratings in the $y$ direction, merging all the initial propagation paths in the $x$ direction with a stable interferometric network, and filtering out the unwanted diffraction orders at the output plane. The automated feature is attained by utilizing a single programmable phase-only SLM, where properly designed diffraction gratings displayed on its screen will implement the desired transformations. We discuss the losses in the process which is, in general, inherently nonunitary, present some examples of transformations and analyze how they will be affected by the pixelated structure of the SLM screen. 

This paper is organized as follows: in Sec.~\ref{sec:Basic} we briefly review the formalism of quantum operations and define the multipath qudit states encoded in a 1D spatial domain. In Sec.~\ref{sec:MainResults} we describe our proposal for automated transformations on these qudits. In Sec.~\ref{sec:Examples} we present some examples of possible transformations, discuss how to implement them in practice, and analyze the effects of the pixelation in the SLM. Finally, in Sec.~\ref{sec:Conclusion} we conclude and discuss some perspectives.

\section{\label{sec:Basic} Basic concepts}

\subsection{Quantum operations}

Mathematically, operations in quantum states are represented by a linear, completely positive, and trace non-increasing map $\Lambda$ \cite{NIELSEN,Kraus83} transforming a quantum state $\rho$  into another quantum state $\rho'$, i.e.,
\begin{equation}
\rho\rightarrow\rho'=\Lambda(\rho),
\end{equation} 
where $\rho$ and $\rho'$ act on a $D$- and $d$-dimensional Hilbert space ($\mathcal{H}$ and $\mathcal{H}'$) respectively. The map $\Lambda$ always admits a decomposition called Kraus representation, in which
\begin{equation}
 \Lambda (\rho) = \sum_l  K_l\rho K_l^{\dagger},
 \label{Kraus}
\end{equation}
where $\{K_l\}$ is the set of Kraus operators. For a given $\Lambda$, a set of Kraus operators can always be found with at most $Dd$ elements \cite{Kraus83} . They satisfy $\sum_l K_l K_l^{\dagger}\leq \mathbb{I}$, where $\mathbb{I}$ is the identity acting on $\mathcal{H}$ and the equality holds for trace-preserving maps. 

In the presentation of our proposal in Sec.~\ref{sec:MainResults}, we focus our discussion into transformations ($\mathcal{M}$), both unitary and nonunitary, taking a pure state into another pure state, i.e.,
\begin{equation}
|\psi\rangle\rightarrow|\psi'\rangle=\mathcal{M}|\psi\rangle.
\end{equation}
In Sec.~\ref{sec:gen_maps} we shall extend it to more general transformations taking a pure state to a mixed one.

\subsection{Single-photon multipath encoded qudits}

Let us consider a paraxial and monochromatic single-photon multimode field propagating along the $z$ direction. Assuming purity, we can write its state in a given transverse plane $z$ as \cite{LEOFENDAS,Monken98,Walborn2010}:
\begin{equation}
\ket{\Psi_z} = \int d\mathbf{x}\, \psi_z(\mathbf{x}) \ket{1\mathbf{x}},
\label{fieldToState}
\end{equation}
where $\mathbf{x} = (x,y)$ is the transverse position coordinate and $\psi_z(\mathbf{x})$ is the field amplitude profile at this plane  satisfying $\int d\mathbf{x}|\psi_z(\mathbf{x})|^2=1$. Now, let $\psi_z(\mathbf{x})$ be given by a superposition of Gaussian functions of radius $\omega_z$ centered at $y=0$ and $x=x_l$ (for $l=1,\ldots,D$), and modulated by complex coefficients $\alpha_l$. Thus
\begin{align}
\psi_z(\mathbf{x})&=\sum_{l=1}^{D}\alpha_lA\exp{\left[-\frac{(x-x_l)^2+y^2}{\omega_z^2}\right]},
\label{eq:Gauss_superp}
\end{align} 
where $A$ is a normalization factor and $\sum_l|\alpha_l|^2=1$. We define the $l$-th Gaussian mode displaced in the $x$ direction as
\begin{align}
\ket{\mathcal{X}_l^z} &\equiv \int d\mathbf{x}\,A\exp{\left[-\frac{(x-x_l)^2+y^2}{\omega_z^2}\right]}\ket{1\mathbf{x}}
\nonumber\\
&=\int d\mathbf{x}\,G_z(x-x_l)G_z(y)\ket{1\mathbf{x}},
\label{Gauss}
\end{align}
where $G_z(\xi)=\sqrt{A}\exp{(-\xi^2/\omega_z^2)}$. Replacing Eqs.~(\ref{eq:Gauss_superp}) and (\ref{Gauss}) into Eq.~(\ref{fieldToState}), we finally arrive at
\begin{equation}
\ket{\Psi_z} = \sum_{l=1}^{D}\alpha_l\ket{\mathcal{X}_l^z}.
\label{eq:multipath_state}
\end{equation}
If $x_l=(l-1)\chi$ in Eq.~(\ref{eq:Gauss_superp}) and $\chi>2\omega_z$, the overlap between the Gaussians will be negligible and, with good approximation, we will have
\begin{align}
\langle \mathcal{X}_l^z|\mathcal{X}_j^z\rangle \approx \delta_{lj},
\label{orthogonality}
\end{align}
so that the states $\{|\mathcal{X}_l^z\rangle\}_{l=1}^{D}$ will be nearly mutually orthogonal. Therefore, the single-photon state in Eq.~(\ref{eq:multipath_state}) will represent a multipath qudit with the information encoded in the paths defined by the Gaussian modes (\ref{Gauss}) in a 1D spatial domain. This state could be prepared, for instance, by sending a single photon with a collimated Gaussian profile $\psi_z(\mathbf{x})$ through a properly designed set of wave-plates and polarizing beam displacers, as recently demonstrated in Refs.~\cite{TESEELIAS,ADANELIAS,CHINESES,UnitaryZeilinger,Hu2016}.

\section{Transformations on multipath qudit states}
\label{sec:MainResults}

We now describe our proposal to implement a given operation represented by the matrix $\mathcal{M}$ that transform the multipath qudit state $|\Psi_z\rangle$ in Eq.~(\ref{eq:multipath_state}) in the following way
\begin{equation}
\mathcal{M}:|\Psi_z\rangle\;\rightarrow\;|\Psi'_{z'}\rangle=\sum_{l=1}^{d}\beta_l|\mathcal{Y}_l^{z'}\rangle,
\label{eq:proposed_transform}
\end{equation}
where $\sum_l|\beta_l|^2=1$, $d$  is an arbitrary positive integer, and $|\mathcal{Y}_l^{z'}\rangle$ is given by Eq.~(\ref{Gauss}) by interchanging $x$ and $y$. Thus, the qudit state initially encoded as a superposition of Gaussian modes displaced along the $x$ direction at an input transverse plane $z$, will be transformed into another qudit state given by the superposition of the Gaussian modes along the $y$ direction at an output transverse plane $z'$.

\subsection{Phase modulation in the $y$ direction}

\begin{figure}[tbp]
\begin{center}
 \includegraphics[width=.3\textwidth]{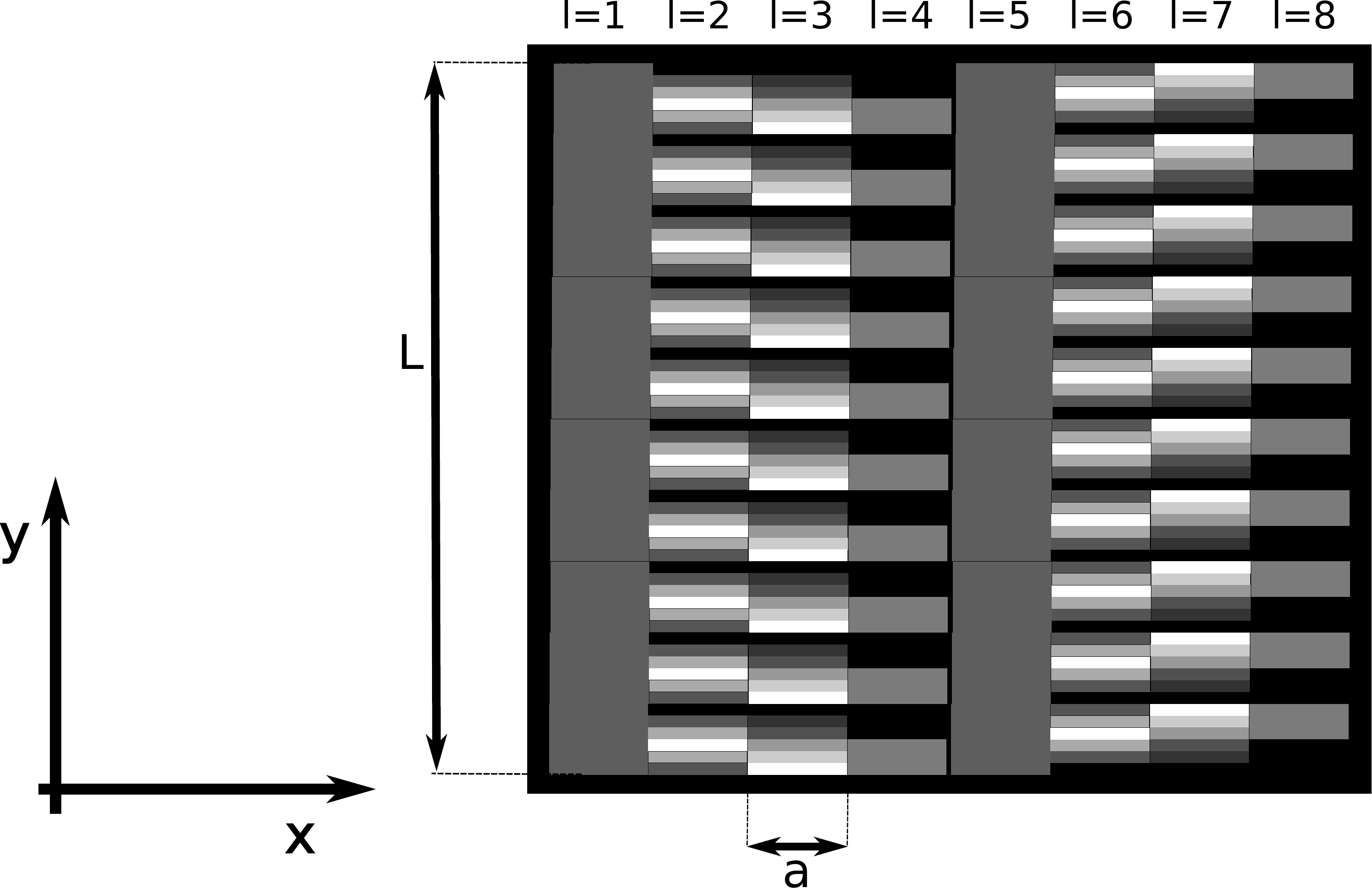}
\caption{Example of a phase mask addressed at the SLM that
we will consider in this work. Its transmission function is
given by Eq. \eqref{eq:transmission_func} for $D = 8$, where the $\Phi_l(y)$\textsuperscript{'}s are diffraction
gratings.}
\label{SLMscreen}
\end{center}
\end{figure}

The first key element in our proposal is the use of a phase-only SLM acting on the $y$ direction of the single-photon field profile. Consider a rectangular SLM divided into $D$ non-overlapping rectangular regions of width $a>2\omega_z$ and centered at $x_l=(l-1)\chi$ for $l=1,\ldots,D$. At each region we address a given function $\Phi_l(y)$ to be specified later.
Thus the transmission function of this SLM can be written as
\begin{equation}
T(\mathbf{x})=\sum_{l=1}^{D}e^{i\Phi_l(y)}{\rm Rect}\left(\frac{x-(l-1)\chi}{a}\right){\rm Rect}\left(\frac{y}{L}\right),
\label{eq:transmission_func}
\end{equation}
where $L$ is the size of the SLM in the $y$ direction. Figure~\ref{SLMscreen} shows an example of a phase mask we shall consider here. The SLM acts according with the polarization state of the incoming field. We assume that the photon is horizontally polarized which will be the working direction of the SLM. With these assumptions, each Gaussian mode $l$ is modulated only by the phase function $e^{i\Phi_l(y)}$. Therefore, using Eqs.~(\ref{eq:Gauss_superp}) and (\ref{eq:transmission_func}), the transmitted field profile $\psi_{z_T}(\mathbf{x})$ will be given by 
\begin{align}
\psi_{z_T}(\mathbf{x})&=T(\mathbf{x})\psi_z(\mathbf{x}) \nonumber\\
&\approx\sum_{l=1}^{D}\alpha_lG_z(x-x_l)G_z(y)e^{i\Phi_l(y)}.
\end{align}
To set the notation for the next discussion, the photon state in a given transverse plane $z_1$ after transmission through the SLM, will be
\begin{equation}
\sum_{l=1}^{D}\alpha_l\int d\mathbf{x}\,G_{z_1}(x-x_l)\mathcal{P}_{z_1}\{G_z(y)e^{i\Phi_l(y)}\}|H\rangle\ket{1\mathbf{x}},
\label{eq:state_z1}
\end{equation}
where $|H\rangle$ is the state of horizontal polarization, that will be explicitly shown in the following calculations; $\mathcal{P}_{z_1}\{\cdot\}$ describes the propagation of the modulated $y$ component of the field profile by a distance $z_1$. Its actual effect will be described later. The $x$ component of the profile, which is not modulated by the SLM, propagates without changing its shape, since we have considered it to be collimated. 

\subsection{Merging the paths in the $x$ direction}

\begin{figure}[tbp]
\begin{center}
 \includegraphics[width=.48\textwidth]{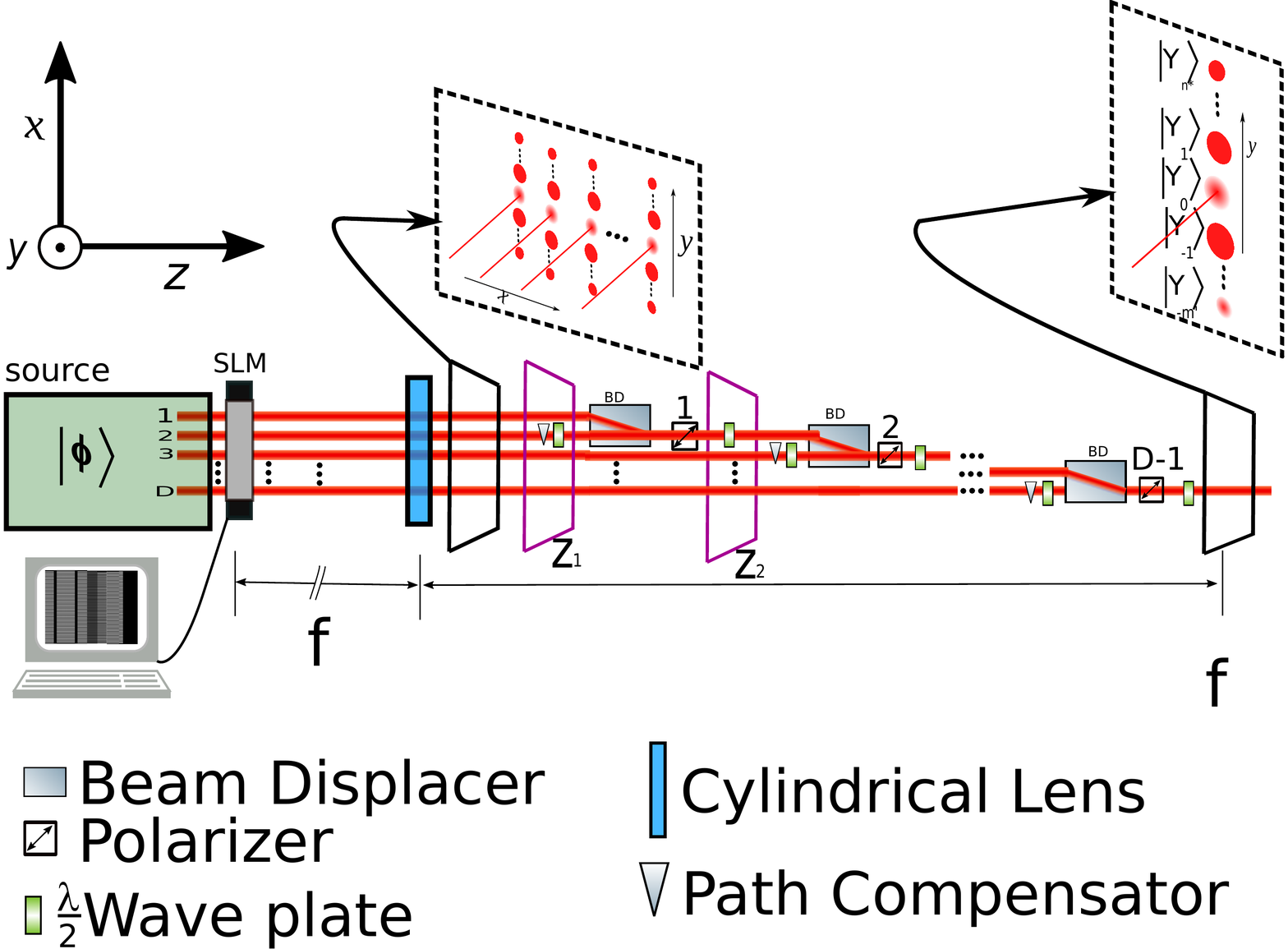}
\caption{Sketch of the proposed setup for implementing automated state transformations on single-photon multipath qudit states. A source generates such states encoded in the $x$ direction [see Eq.~(\ref{eq:multipath_state})] with horizontal polarization. A programmable phase-only SLM addressed by a phase mask given by (\ref{eq:transmission_func}) modulates the $y$ component of the field profile, transforming the qudit state into (\ref{eq:state_z1}). The first inset shows, with purposely exaggeration, a representation of the state at this point. The interferometric network described in the text accomplishes the transfer of the encoding from $x$ to $y$ direction, generating the state (\ref{eq:Psi_out_interf}). The cylindrical lens, which has the SLM in its focus, ensures that the diffraction orders will not separate too much and have approximately parallel propagation along the interferometer. It also ensures that they will be described at the back focal plane by the Fourier coefficients of the gratings displayed at the SLM screen. The transformations are defined by the phase gratings displayed at the SLM screen, as we show in Sec.~\ref{Subsec: QDits} and produce the unfiltered state \eqref{estadoFinal}. After a proper spatial filtering, the transformation \eqref{eq:proposed_transform} is accomplished, and the transformed multipath qudit state is now encoded in the $y$ direction, as illustrated in the second inset.}
\label{setup}
\end{center}
\end{figure}

The second key element of this proposal is to merge the $D$ paths in the $x$ direction (which define the qudit state) into a single one. For this we shall use the photon polarization as an auxiliary system and the interferometric arrangement similar to the ones in \cite{OBrien2003Interf,WhiteInterf,Hu2016} sketched in Fig.~\ref{setup}. This interferometer is composed by $D-1$ polarizing beam displacers (PBD), polarizers and optical path compensators, and $2D-2$  half-wave plates (HWP). In the PBDs we assume that a vertically polarized photon is directly transmitted and a horizontally polarized undergoes a lateral displacement in the $x$ direction by a distance $\chi$ equal to the center-to-center separation between neighbor paths. For each PBD, the HWPs before and after it are set to transform the incoming polarization state into $|V\rangle$ and $|H\rangle$, respectively. The optical compensators provide the superposition for the transmitted and displaced paths while the polarizers after the PBDs are used to erase the which-path information, thus ensuring the required interference effect. 

Let us see now how the multipath qudit state after the action of the SLM [Eq.~(\ref{eq:state_z1})] evolves along this setup. We start the procedure for the paths 1 and 2 in Fig.~\ref{setup}. In the path 2, the HWP transforms  $|H\rangle\rightarrow|V\rangle$. After the PBD and the compensator, the paths 1 and 2 are merged and $G_{z_1}(x-x_1)\rightarrow G_{z_1}(x-x_2)$. Polarizer 1 projects the photon polarization in this merged path into $\frac{1}{\sqrt{2}}(|H\rangle+|V\rangle)$ and a HWP transform it back to $|H\rangle$. After all this, the state postselected from the polarization projection becomes  
\begin{align}
\frac{1}{\sqrt{2}}\sum_{l=1}^{2}\alpha_l\int d\mathbf{x}\,G_{z_2}(x-x_2)\mathcal{P}_{z_2}\{G_z(y)e^{i\Phi_l(y)}\}|H\rangle\ket{1\mathbf{x}}
\nonumber\\
+ \sum_{l=3}^{D}\alpha_l\int d\mathbf{x}\,G_{z_2}(x-x_l)\mathcal{P}_{z_2}\{G_z(y)e^{i\Phi_l(y)}\}|H\rangle\ket{1\mathbf{x}},
\end{align}
where the first term corresponds to the merged paths $1+2\rightarrow 1'$ and the second to the remaining $D-2$ paths. This procedure is iterative: for $p=2,\ldots,D$, the polarization in the path $p$ is transformed as $|H\rangle\rightarrow|V\rangle$; then this path is merged with path $p-1$ (which may result from previous mergings) in the $(p-1)$-th PBD. In order to erase the which-path information and minimize losses, the $(p-1)$-th polarizer projects the photon polarization into $\frac{1}{\sqrt{p}}(\sqrt{p-1}|H\rangle+|V\rangle)$. Finally, this polarization is rotated to $|H\rangle$. The general state postselected after these transformations can be written as
\begin{align}
\frac{1}{\sqrt{p}}\sum_{l=1}^{p}\alpha_l\int d\mathbf{x}\,G_{z_{p}}(x-x_p)\mathcal{P}_{z_{p}}\{G_z(y)e^{i\Phi_l(y)}\}|H\rangle\ket{1\mathbf{x}}
\nonumber\\
+ \sum_{l=p+1}^{D}\alpha_l\int d\mathbf{x}\,G_{z_{p}}(x-x_l)\mathcal{P}_{z_{p}}\{G_z(y)e^{i\Phi_l(y)}\}|H\rangle\ket{1\mathbf{x}},
\end{align}
now, with the first (second) term corresponding to the merged $p$ paths (remaining $D-p$ paths).

Therefore, at the output of the interferometer in Fig.~\ref{setup}, i.e., for $p=D$ in the above description, the postselected state of the output photon will be transformed as
\begin{align}
|\Psi_z\rangle\;\rightarrow\; |\Psi_{z_D}'\rangle &\equiv\sum_{l=1}^{D}\alpha_l'|\mathcal{W}_l^{z_D}\rangle,
\label{eq:Psi_out_interf}
\end{align}
where $\alpha'_l=\alpha_l/\sqrt{D}$ and
\begin{equation}
|\mathcal{W}_l^{z_D}\rangle=\int d\mathbf{x}\,G_{z_{D}}(x)\mathcal{P}_{z_{D}}\{G_z(y)e^{i\Phi_l(y)}\}\ket{1\mathbf{x}}.
\label{eq:W_state}
\end{equation}
In the above equation, we dropped the polarization state which will play no role from now on and redefined the origin in the $x$ direction making $x_D=0$. Note that the spatial mode label $l$ is now only in the $y$ component of the single-photon field profile. Thus, with the whole procedure described so far we were able to transfer the encoding from a one-dimensional transverse spatial domain, $x$, to the orthogonal one, $y$. However, in order to implement the proposed transformation (\ref{eq:proposed_transform}), we must be capable of transforming these $D$ modes $|\mathcal{W}_l^{z_D}\rangle$ defined in the $y$ direction into a set of $d$ orthogonal modes, which is our next topic.

As we will see next, the $\Phi_l(y)$ functions in the SLM modulating the $y$ component of the field profile,  will be phase gratings. Thus, for the interferometer to work properly as we described, the generated diffraction orders in $y$ direction must propagate in parallel and without separating too much (in order to pass through the optical elements). To achieve this, we consider the SLM to be in the focus of a cylindrical lens as shown in Fig.~\ref{setup}.

\subsection{\label{Subsec: QDits} Creating orthogonal modes in the $y$ direction}

To create the required orthogonal modes, we first define the back focal plane of the cylindrical lens as our plane of observation after merging the $D$ paths, i.e., $z_D=f$, as shown in Fig.~\ref{setup}. In this case, we replace $\mathcal{P}\{\cdot\} \rightarrow \mathcal{F}\{\cdot\}$ in Eq.~(\ref{eq:W_state}), where $\mathcal{F}\{\cdot\}$ represents the Fourier transform of the field profile in $y$ direction \cite{GOODMAN}. At the focal plane we have $k_y=ky/f$, where $k_y$ is the $y$ component of the photon wave vector and $k$ is its wavenumber. Thus, we can denote the $l$-th $y$-component mode function to be computed, as
\begin{equation}
\mathcal{G}_l(y)=\mathcal{F}\{G_z(y')e^{i\Phi_l(y')}\}(k_y).
\label{eq:G_y}
\end{equation}
If the $\Phi_l(y')$'s are periodic functions with period $T$, we can expand the Fourier series
\begin{equation}
e^{i\Phi_l(y')}=\sum_{j=-\infty}^{\infty}C_{jl}e^{2\pi ijy'/T},
\end{equation} 
where $C_{jl}$ are the corresponding Fourier coefficients $C_{jl}=\frac{1}{T}\int_{0}^{T}dy'\,e^{i\Phi_l(y')}e^{-2\pi ijy'/T}$. Therefore, Eq.~(\ref{eq:G_y}) becomes
\begin{align}
\mathcal{G}_l(y) &= \sum_{j=-\infty}^{\infty}C_{jl}\mathcal{F}\{G_z(y')e^{2\pi ijy'/T}\}(k_y) \nonumber\\
&=\sum_{j=-\infty}^{\infty}C_{jl}\mathcal{F}\{G_z(y')\}(k_y-2\pi j/T) \nonumber\\
&\equiv\sum_{j=-\infty}^{\infty}C_{jl}\tilde{G}_f\left[\frac{k}{f}(y-y_j)\right],
\end{align}
where $\tilde{G}_f=\mathcal{F}\{G_z\}$ and $y_j=2\pi jf/Tk$. From Eqs.~(\ref{eq:Psi_out_interf}) and (\ref{eq:W_state}), the single-photon field profile at the focal plane of the lens, $\psi'_f(\mathbf{x})$, will be written as\begin{align}
 \psi'_f(\mathbf{x}) & = 
\sum_{l=1}^D\alpha'_l\sum_{j=-\infty}^{\infty}C_{jl}G_f(x)\tilde{G}_f\left[\frac{k}{f}(y-y_j)\right]
\nonumber\\
&\equiv \sum_{j=-\infty}^{\infty}\beta'_{j}G_f(x)\tilde{G}_f\left[\frac{k}{f}(y-y_j)\right],
\label{FieldzF}
\end{align}
where, recalling that $\alpha'_l=\alpha_l/\sqrt{D}$,
\begin{equation}
\beta'_j=\frac{1}{\sqrt{D}}\sum_{l=1}^{D}\alpha_lC_{jl}. 
\label{eq:beta_1}
\end{equation}
Since we considered the qudits encoded into Gaussian spatial modes [see Eq.~(\ref{eq:Gauss_superp})], the Fourier transform $\tilde{G}_f\left[(y-y_j)k/f\right]$ are also Gaussian with radius $\omega'_f=2f/\omega_zk$. This shows the importance of the path components having, as transverse profiles, eigenstates of the Fourier Transform. The field profile will then be a superposition of Gaussian modes along the $y$ directions separated by a distance $\Delta y = |y_j - y_{j+1}| = 2\pi f/Tk$. If the gratings period satisfies $T<\pi\omega_z/2$, the overlap between these Gaussian modes will be negligible and, similarly to Eq.~\eqref{orthogonality}, we can write
\begin{equation}
\langle\mathcal{Y}_l^f|\mathcal{Y}_j^f\rangle\approx \delta_{lj},
\label{eq:ortog_Y}
\end{equation}
where 
\begin{equation}
|\mathcal{Y}_j^f\rangle=\int d\mathbf{x}\,G_f(x)\tilde{G}_f\left[\frac{k}{f}(y-y_j)\right]|1\mathbf{x}\rangle.
\label{eq:Y_mode}
\end{equation}
Therefore, using Eqs.~(\ref{FieldzF}), (\ref{eq:beta_1}), and (\ref{eq:Y_mode}), the single-photon state postselected from the interferometer---after the action of the SLM---will be given by
\begin{align}
 \ket{\Psi'_f} &= \frac{1}{\sqrt{D}}\sum_{l=1}^{D}\sum_{j=-\infty}^{\infty}\alpha_lC_{jl}\ket{\mathcal{Y}_j^f}
\nonumber\\
&= \sum_{j=-\infty}^{\infty}\beta'_j\ket{\mathcal{Y}_j^f},
\label{estadoFinal}
\end{align}
with the Gaussian modes $|\mathcal{Y}_j^f\rangle$'s satisfying the orthogonality relation (\ref{eq:ortog_Y}) and $\sum_j|\beta'_j|^2=1$. 

Comparing Eqs.~\eqref{eq:proposed_transform} and \eqref{estadoFinal}, one can see that the state transformation $|\Psi_z\rangle\rightarrow|\Psi'_f\rangle$ is determined by the Fourier coefficients $\{C_{jl}\}$ describing the phase diffraction gratings addressed at the SLM. More specifically, each coefficient $\beta'_j$ [see Eq.~(\ref{eq:beta_1})] of the transformed state is constructed by multiplying each initial state coefficient $\alpha_l$ by the $j$-th Fourier coefficient from the $l$-th diffraction grating, $C_{jl}$, and adding them together. This means that in order to have a transformation represented by the matrix $\mathcal{M}$ it is necessary to choose the right gratings to have $C_{lj}\propto M_{lj}$: each $\Phi_l$ corresponds to the $l$-th column of the implemented $\mathcal{M}$. Since the gratings may be generated and controlled in automated way in the SLM, the transformations proposed here can be completely automated as well.

\subsection{Spatial Filtering}

In Eq.~(\ref{estadoFinal}), all the orders are taken into account. In order to have finite dimension states and in a more realistic scenario, a spatial filtering is used.

The role of the spatial filtering is to  define a finite Hilbert space of dimension $d$ for the resulting transformed state, truncating the sum in expression \eqref{estadoFinal}. It can be done by filtering out all but $d$ orders, considering the other orders as losses. This can be implemented at plane $z=z_D$ or as soon as the orders can be distinguished (as depicted in the second inset of Fig.~\ref{setup}). For simplicity, we will assum that in either case the same orders are filtered in each path. Let us use the following nomenclature: the orders $j$ that are not filtered belong to the interval $j_{1}\leq j\leq j_{2}$. Then, the fraction of the photons that will be lost by spatial filtering is given by
\begin{equation}
 \tau_{{\mathcal{M}}} = \sum_{l=1}^D \left(\sum_{m=-\infty}^{j_1-1} |C_{lm}|^2 + \sum_{n=j_2+1}^\infty|C_{ln}|^2\right).
\label{filtering}
\end{equation}
It is important to note that this $\tau_{{\mathcal{M}}}$ factor depends
on the matrix ${\mathcal{M}}$ which is implemented and we will see some numeric values for some examples in Sec. \ref{sec:Examples}. 
Finally, Eq. \eqref{estadoFinal} is modified by filtering as follows: 
\begin{align}
\ket{\Psi_f}&=\sum_{l=1}^{D}\sum_{j=j_1}^{j_1+d-1}\frac{\alpha_l}{\sqrt{D}}C_{jl}\ket{\mathcal{Y}_j} 
\nonumber\\
&=\sum_{j'=1}^{d}\beta_{j'}\ket{\mathcal{Y}_{j'}^f},
\label{StateFiltering}
\end{align}
where $j'=j -j_1+1$.

We can see that in this case  ${\mathcal{M}}_{j'l} \propto C_{jl}$, but now considering only the non-filtered orders. This can be written in matrix form as
\begin{widetext}
\begin{align}
&\left[\begin{array}{cccccccc}
m_{11}  &&    m_{12}     &&\ldots && m_{1D} \\
m_{21}  &&    m_{22}     &&\ldots && m_{2D} \\
\vdots      &&    \vdots     &&\vdots && \vdots\\
m_{d1} &&  m_{d2} &&\ldots && m_{dD} \\
\end{array} \right] = 
\frac{1}{\sqrt{D}}\left[\begin{array}{cccccccc}
C_{j_11}     &&    C_{j_12}   &&\ldots && C_{j_1D} \\
C_{(j_1+1)1}     &&    C_{(j_1+1)2}   &&\ldots && C_{(j_1+1)D} \\
\vdots       &&    \vdots     &&\vdots && \vdots  \\
C_{(j_1+d-1)1} &&  C_{(j_1+d-1)2} &&\ldots && C_{(j_1+d-1)D} \\
\end{array} \right].
\label{MeCim}
\end{align}
\end{widetext}

An interesting possibility that can increase the control of the operation done by this proposal is to use another transmission phase-only SLM before the state enters the interferometer, for example the plane $z=z_1$ in Fig.~\ref{setup}. If the orders are already spatially separated so one
can act on each order individually, this second SLM can change the phase of each coefficient, changing the matrix ${\mathcal{M}}$. In this case it can also allow to correct phase factors that can arise from the propagation of higher orders through the PBD.

\section{\label{sec:Examples} Examples and techniques}

\subsection{Some phase gratings and their operation}

As seen in the previous section, the phase gratings configuration at the SLM' screen determine
the operation that is implemented in the initial state: each phase grating $\Phi_l(y')$ compose, by
its Fourier coefficients, the $l$-th column of the matrix ${\mathcal{M}}_{lj}$. Now we show
how to use some simple phase gratings in order to make important operations, such as permutation and projections. We will present a study of their coefficients as well as examples of ${\mathcal{M}}$ that can be achieved with these gratings. Most of these examples will be in
Hilbert spaces of dimension $d=D=3$ for the initial and final qudit, selecting orders $-1\leq j\leq 1$, but this is not an intrinsic limitation of this proposal, as seen in Sec. \ref{sec:MainResults}.  

\subsubsection{Saw-tooth grating} 

\begin{figure}[h!]
 \includegraphics[width = 8.8cm, height=3.5cm]{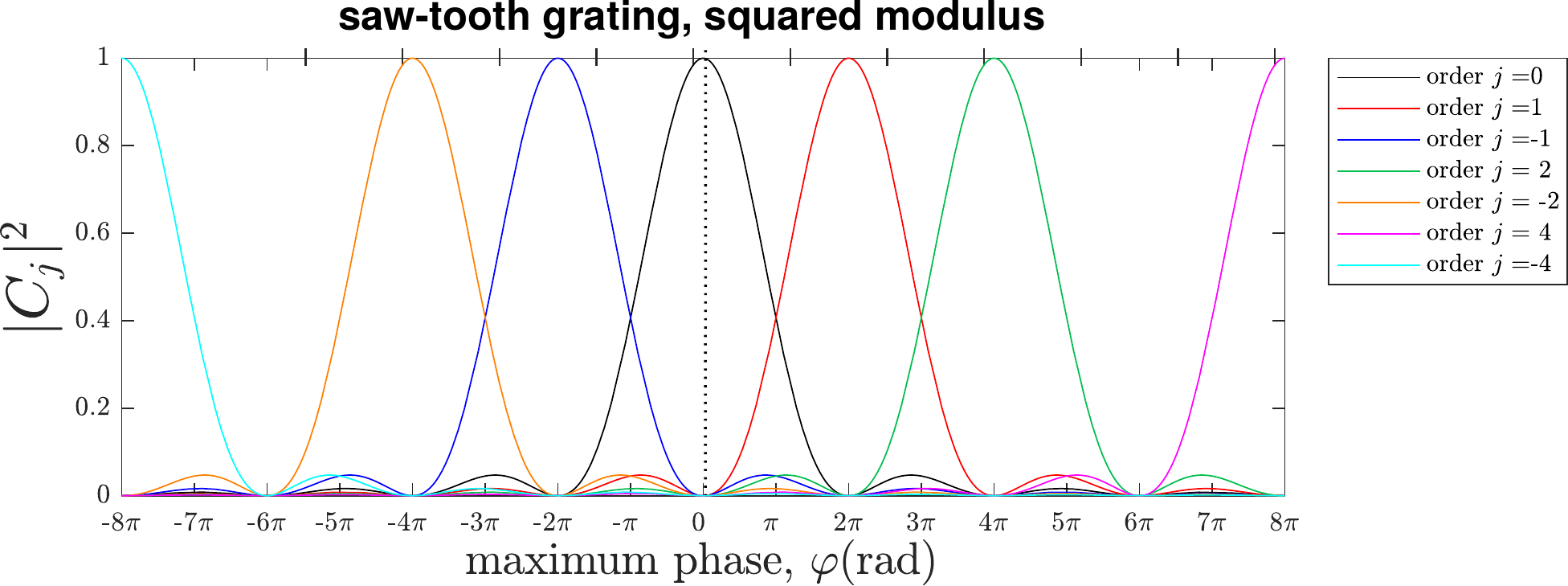}
\caption{The behaviour of the modulus squared of the coeficients of the saw-tooth grating,
as function of the maximum phase value $\varphi$. As it is possible to get up to $\varphi \approx 8\pi$ modulation
with a SLM, this is the maximum choose in the graphic. The values of $\varphi<0$ represent an descending saw-tooth grating and the $\varphi>0$ region represents an ascending saw-tooth grating (see \eqref{CmLInear}). The dotted line at $\varphi = 0$ represents a division between the these two regions. The negative values of $\varphi$ means a saw-tooth grating that 
varies in the opposite direction. The orders $j=1,0,-1,2,-2,4,-4$ are showed. This grating has the interesting
characteristic of having $C_{\bar{m}} = 1$ for $\varphi = \bar{m}2\pi$.}
\label{Lin_mod}
\end{figure}

\begin{figure}[h!]
 \includegraphics[width=8.8cm, height=3.5cm]{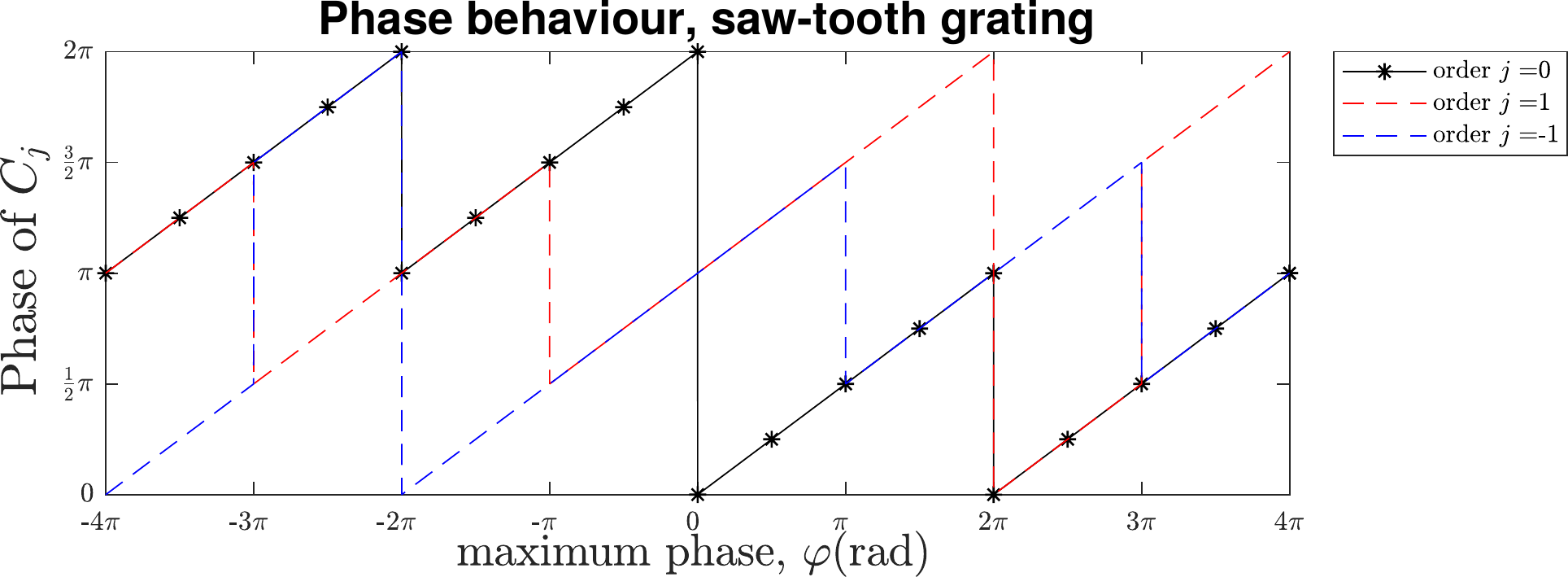}
\caption{The behaviour of the relative phase between the coefficients of the saw-tooth grating and the incoming path component. The $\varphi$ with negative sign represents the descending saw-tooth grating while the positive sign representing an ascending saw-tooth grating. The interval $\{-4\pi,4\pi\}$ shows a typical behaviour for the orders $j=\pm1$ and $j=0$. . All orders have a linear dependence on $\varphi$ with periodically $\pi$ phases added by the change of signal of the $\sinc$ function. The plot is made modulo $2\pi$. }
\label{Lin_phase}
\end{figure}

The first one to be considered is the saw-tooth phase-grating, determined by:
\begin{align}
 \Phi(y) &= \frac{\varphi y}{T},
\end{align}
where $\varphi$ is the maximum phase value of the grating.
The coefficientes of the Fourier expansion for this idealized continuous grating is given by the following expressions:
 \begin{align}
C_0 &= e^{i\varphi/2}\sinc(\varphi/2),\\[2mm]
C_j &= e^{i(\varphi/2-j\pi)}\sinc(\varphi/2-j\pi).
 \label{CmLInear}
 \end{align}
The behavior of this $C_j$'s, in function of the value of $\varphi$ is showed in Fig. \ref{Lin_mod} and \ref{Lin_phase}.

It is important to note that for each value of $\varphi$, a different operation is implemented in the component associated to the grating. For  $\varphi = \bar{m}2\pi$, all the coefficients are null except for the $C_{\bar{m}}$ coefficient, which is equal to $1$.
This means that if such a phase grating is used at the \textit{i}-th path, its contribution to the final state will be only to the
$\bar{m}$-th component of the final state. For example, if the initial state is given by:
\begin{equation}
\ket{\mathcal{X}_0} = \begin{pmatrix} 0 \\ 1 \\0\end{pmatrix}_{X},
\label{X_0}
\end{equation}
and the saw-tooth phase grating with $\varphi = 2\pi$ is used, the final state will be, after renormalization:
\begin{equation}
\ket{\mathcal{Y}_1} = \begin{pmatrix} 1 \\ 0 \\0\end{pmatrix}_{Y}.
\label{Y_1}
\end{equation}

With such a phase grating it is possible to make Right-, Left- and permutation operations \cite{NIELSEN}. The ${\mathcal{M}}$ matrix for the Left- operation in a $3$ dimensional Hilbert space, for example, is:
\begin{equation}
{\mathcal{M}}_{\rm Left} \propto \left(
\begin{array}{ccc}
0 & 0 & 1\\
1 & 0 & 0\\
0 & 1 & 0
\end{array} \right),
\label{LeftOperation}
\end{equation}
and can be performed by our proposal by making $\Phi_3(y') = -\Phi_2(y') = 2\pi/T$ and $\Phi_1(y')=0$.
All these transformations (lowering, raising and permutations) are restrained by the maximum modulation of the SLM. As nowadays SLMs can reach a maximum modulation of approximately $8\pi$, it means that this proposal can be used to get transformations such as \eqref{LeftOperation} up to $d=8$, with current technology. However, this limitation can be overcome
by the development of SLMs with higher maximum modulation.

Another important feature that is achievable with this grating, together with spatial filtering, is to block a path state component: this is made by using a saw-tooth grating with $\varphi = 2\bar{m}\pi$ with a high value of $|\bar{m}|$. In this case
the only non-zero coefficient is filtered out of the setup. This can also be done by making the period of the grating
shorter, in order to increase the separation $\Delta y$ between the orders, reaching the same effect.
With this second possibility, projection into the computational basis is straightforward.
In cases where only permutations or computational basis projections are used, the polarizers can be 
taken out of the setup since there will be no superpositions of different paths in the \textit{x}-direction, making the
expression in \eqref{LeftOperation} an equality, saving losses.

\subsubsection{Binary phase grating} 

\begin{figure}[h!]
 \includegraphics[width = 8.8cm, height=3.5cm]{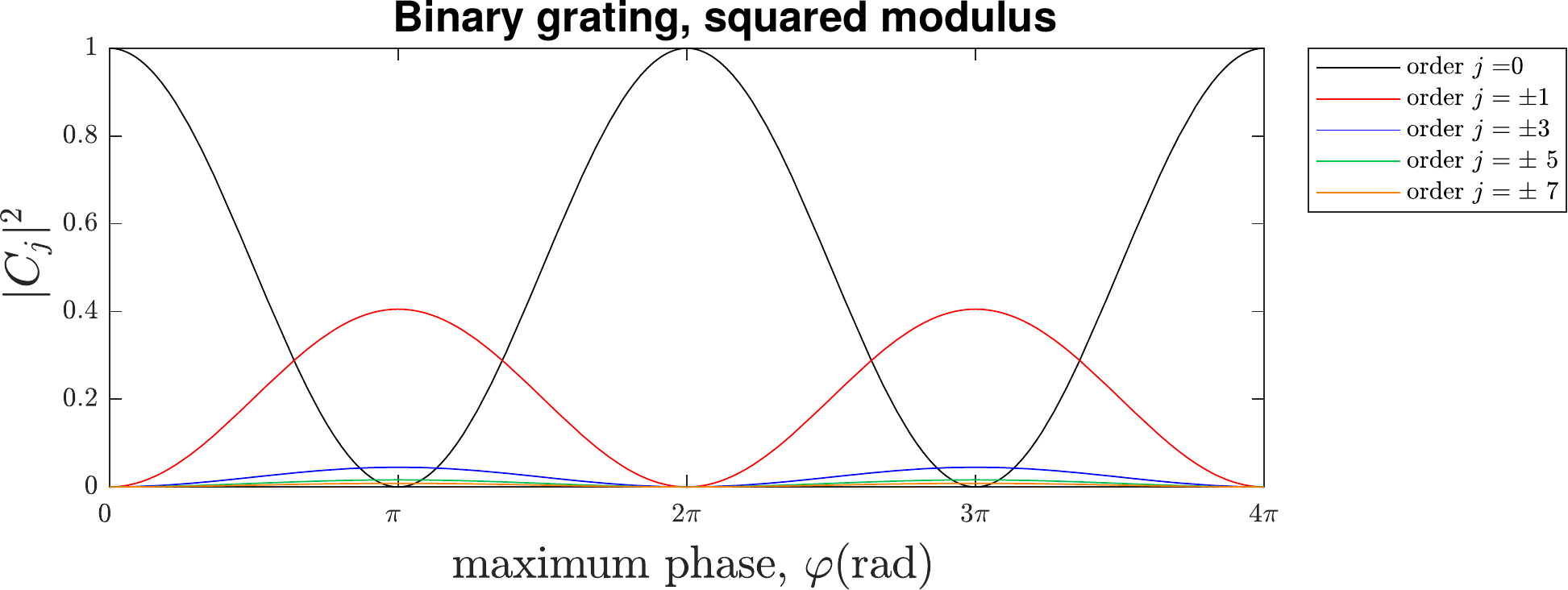}
\caption{The behaviour of the modulus squared of the coefficients of the binary grating, in function of $\varphi$, for the first nine orders.
This grating's coefficients are periodic, with period $2\pi$. At $\varphi = \pi$, the order $j=0$ is null and all the other orders reach their maxima. The coefficients
of $j<0$ have the same squared modulus of the $j>0$ coefficients.}
\label{Bin_mod}
\end{figure}

\begin{figure}[h!]
 \includegraphics[width = 8.8cm, height=3.5cm]{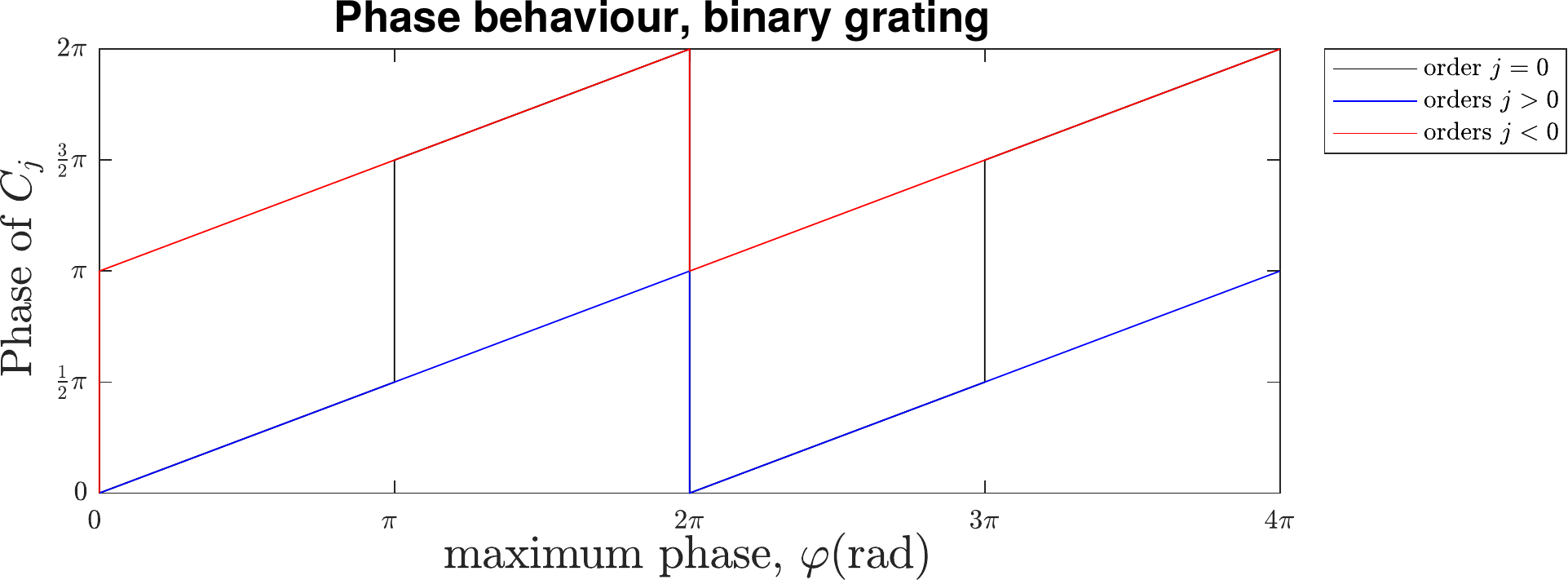}
\caption{The phases of the coefficients calculated from the binary grating. The phase of the orders relative to the incoming state is linear with $\varphi$.
It is interesting that for each value of $\varphi$ that the $j=0$ order goes to $C_0 =0$, it changes by an amount of $\pi$. The plot is made modulo $2\pi$ and shows two periods of the phases.}
\label{Bin_phase}
\end{figure}

Another simple phase grating to use is the binary phase grating represented by the function:
\begin{eqnarray}
 \Phi(y)& = 
\begin{cases}
 0 , & 0\leq y<T/2,\\
\varphi, & T/2\leq y < T,
\end{cases}
\end{eqnarray}
with Fourier coefficients given by:
\begin{eqnarray}
C_0 &=&e^{i\frac{\varphi}{2}}\cos (\varphi/2),\\
C_j &=& 
\begin{cases}
\frac{2}{j\pi}e^{i\frac{\varphi}{2}}\sin(\varphi/2),\; &j\; {\rm odd},\\
 0,\; & m\; {\rm even}.
\end{cases}
\label{CmBin}
\end{eqnarray}
With this phase grating other transformations are possible. It is important to note that all the coefficients with $j<0$ have an opposite phase to the $j>0$ while $|C_j| = C_{-j}$. The central order (determined by $C_0$) have its phase equal to the $j>0$ coefficientes
for $0\leq \varphi<\pi$ and equal to the $j<0$ orders for $\pi<\varphi<2\pi$.  These phase grating's coefficients have the interesting characteristic of being 
periodic in $\varphi$, with period $2\pi$, and the zeroth order is null at $\varphi = (2k+1)\pi$, with $k$ a positive integer number. These behaviours are depicted in Fig. \ref{Bin_mod} and \ref{Bin_phase}.


Using this grating in a qutrit initial state, with $\varphi = \arctan({\pi/2})$ for components $l=1,2$ and $3$ with addition of a constant phase of value $\pi$ in the $l=3$ region of the SLM, it is possible to make an operation proportional 
to the projection in the state $\ket{v} = (1, 1, -1)^T$--the proportionality, not an equality, occurs because it is necessary to make use of the filtering described in the previous section. In matrix form:
\begin{equation}
{\mathcal{M}} \propto \frac{1}{3}\left(
\begin{array}{rrr}
1 & 1 & -1\\
1 & 1 & -1\\
-1 &-1 &1
\end{array} \right);
\end{equation}
The constant of proportionality is $1 - \tau_{\mathcal{M}} = 0.86$ calculated from \eqref{filtering}. Other qutrit projections can be obtained with the use of this phase grating
and the saw-tooth grating. It is important to remind, when considering different operations, that the constant $\tau_{{\mathcal{M}}}$ depends on the matrix ${\mathcal{M}}$.

\subsubsection{Triangular grating}

\begin{figure}[h!]
 \includegraphics[width = 8.8cm, height=3.5cm]{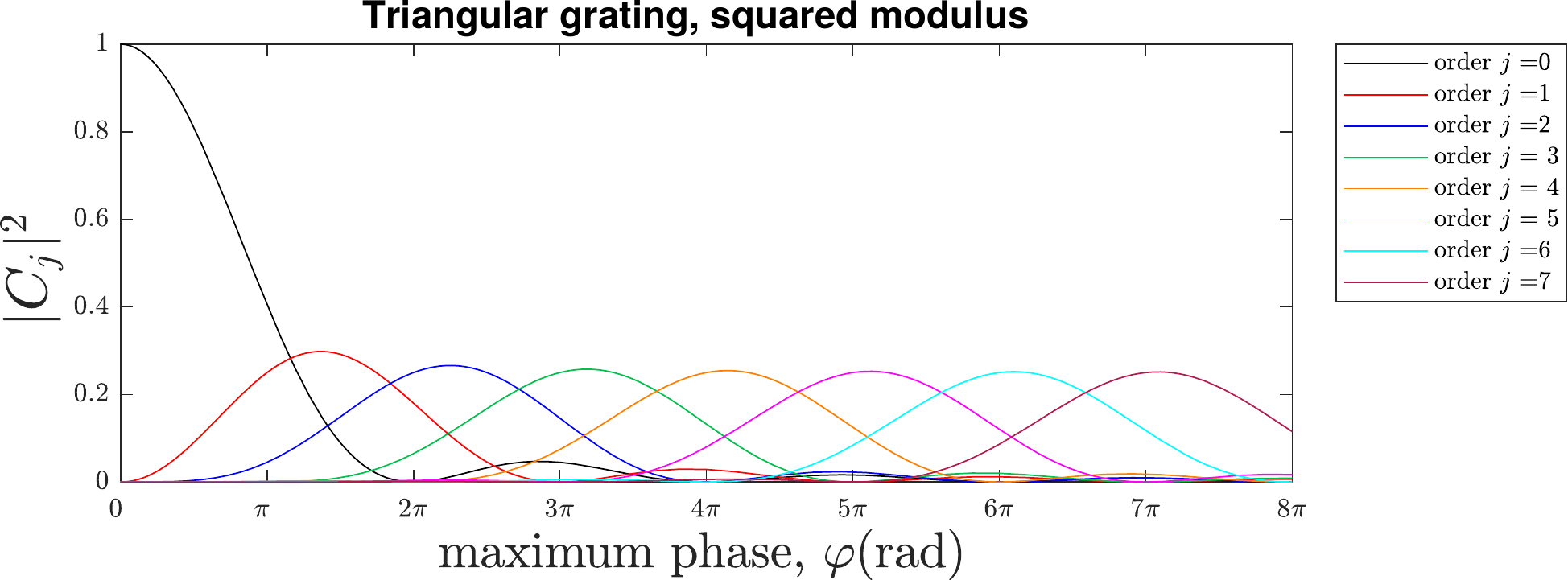}
\caption{The behaviour of the modulus squared of the coefficients of the triangular grating as function of $\varphi$.
As $\varphi$ goes further from zero, higher orders starts to have $|C_j|^2$ not negligible. The maximum for each order, however, is not the same.
At $\varphi = 2\pi$, the order $j=0$ is null. The coefficients
of $j<0$ are identical to the $j>0$ coefficients.}
\label{Triang_mod}
\end{figure}

\begin{figure}[h!]
 \includegraphics[width = 8.8cm, height=3.5cm]{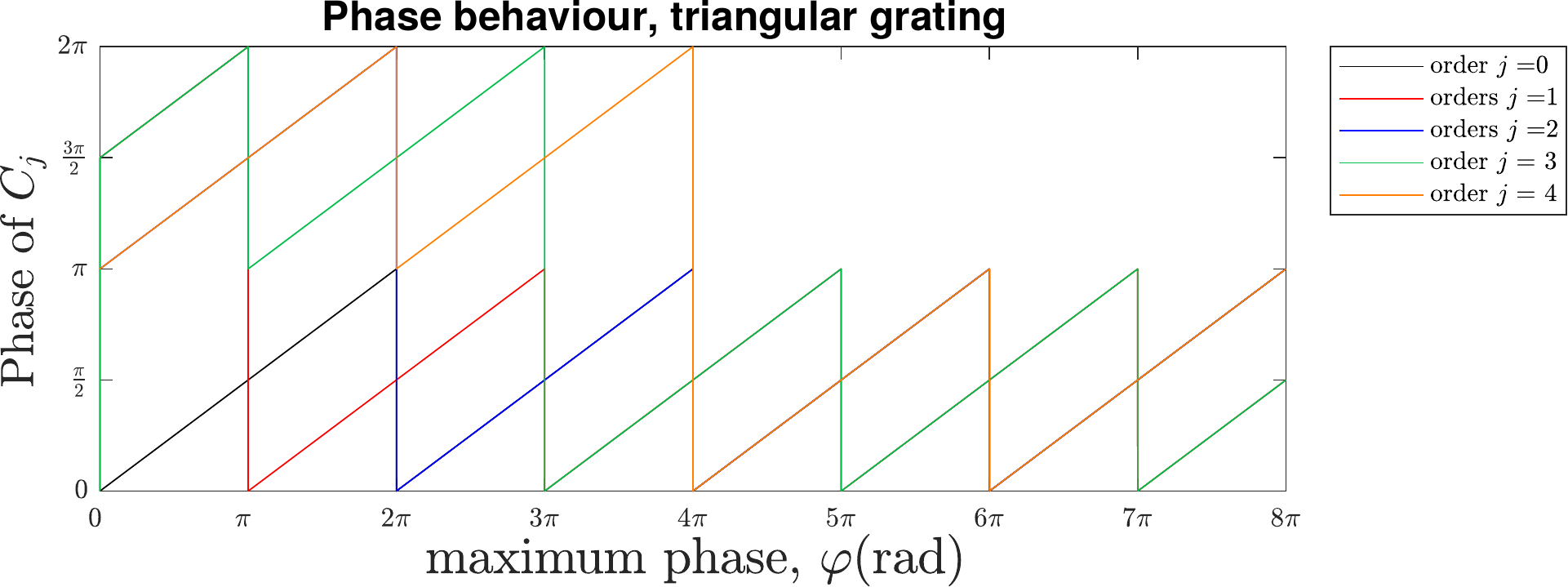}
\caption{The phases of the coefficients for the triangular grating. The phase of the orders relative to the incoming state is linear with $\varphi$. 
The plot is made modulo $2\pi$. Again it is possible to see that the phases have discrete variations of value $\pi$ coming from the $\sinc$ functions.}
\label{Triang_phase}
\end{figure}

This grating is defined as
\begin{align}
\Phi(y) =
\begin{cases}
%\displaystyle
 \frac{-\varphi 2\pi y}{T}, &  -T/2 \leq y \leq 0, \\[2mm]
%\displaystyle
 \frac{\varphi 2\pi y}{T}, & 0 < y \leq T/2,
\end{cases}
\label{Triangular}
 \end{align}
with Fourier coefficients given by
\begin{align}
 C_0 = & -ie^{i\frac{\varphi}{2}}\sinc{(\varphi/2)}, \\
C_j &=  \frac{-i}{2}e^{i\frac{\varphi + \pi j}{2}}\sinc{[(\varphi + \pi j)/2]} \nonumber\\ &-\frac{i}{2} e^{i\frac{\varphi - \pi j}{2}}\sinc{[(\varphi - \pi j)/2]}.
\label{CTriangular}
\end{align}
The behaviour of these coefficients with $\varphi$ is shown in Fig. \ref{Triang_mod} and \ref{Triang_phase}. It is important to note that for this grating
$C_{-j}=C_{j}$, reflecting the simetry of the function shown in Eq. \eqref{Triangular}. 

It is important to mention that this grating also has coefficient values less negligible than the binary grating for higher orders. With the triangular phase function with $\varphi = 2\pi$ in regions $l=1,3$ and the spatial filtering of the initial state component $l=2$ by the proper saw-tooth grating it is possible,
for example, to make the following operation:
\begin{equation}
{\mathcal{M}} \propto \frac{1}{2}\left(
\begin{array}{ccc}
1 & 0 & 1\\
0 & 0 & 0\\
1 & 0 & 1
\end{array} \right),
\end{equation}
which is proportional to the projection at the $\ket{w} = (1,0,1)^T$ state. In this case the normalization constant is $1-\tau_{{\mathcal{M}}} = 0.36$.

\subsection{Combinations, grating displacement and more general maps}
\label{sec:gen_maps}

An important feature of this proposal is that the quantum operations are automated and defined by the 
diffraction gratings used; thereby unbounded possibilities for transformations can be envisaged.
This possibilities, however, are not confined to the different gratings to be used. Combinations of
gratings with coefficients already calculated and use of the same gratings displaced by a distance ``$a$'' can produce new operations. It is also straightforward to produce, with few changes in this setup, convex sum of operations.

\subsubsection{Displacing gratings} 

Different quantum operations can be realized by displacing a phase grating by an integer number $p$ of pixels or an amount $a = p*l$ in the \textit{y}-direction, considering $l$ as the size of the pixel in the \textit{y} direction.
If we write $e^{i\Phi(y')} = f(y')$, and use the superscript $(d)$ for  the displaced function we have:
\begin{align}
C_j^{(d)}&=\int_{-\frac{T}{2}}^{\frac{T}{2}} f^d(y')e^{\frac{-2\pi ij}{T}\,y'}dy' \nonumber\\
&=\int_{-\frac{T}{2}}^{\frac{T}{2}} f(y'-a)e^{\frac{-2\pi ij}{T}y'}dy'\nonumber\\
&= \int_{-\frac{T}{2}-a}^{\frac{T}{2}-a} f(y)e^{\frac{-2\pi ij}{T}(y+a)}dy\nonumber \\
&=e^{\frac{-2\pi ij}{T}a}\int_{-\frac{T}{2}-a}^{\frac{T}{2}-a} f(y)e^{\frac{-2\pi ij}{T}y}dy \nonumber\\
&= e^{\frac{-2\pi ij}{T}a}C_j.
\end{align}

As $a = pl$ and $T = Nl$, we have:
\begin{equation}
C_j^{(d)}\,=\,e^{\frac{-2\pi ij}{N}p}\,C_j,
\end{equation}
where $C_j^{(d)}$ is the coefficient of the displaced grating and $N$ is the total number of pixels in a period $T$.
This shows that by displacing transversally the grating (relative to the position of the center of the gaussian amplitude of the path state component in the \textit{y}-direction) we are able to change the relative phases of the coefficients. It implies that the
same grating with the same $\varphi$ can be used to have different $C_j$'s and, thereby, different operations. This technique used with a saw-tooth grating can be a simple manner to perform on the input state the Pauli matrices in Hilbert spaces of dimension $D=2$, for example.

\subsubsection{Composition of gratings}

Another way of achieving different operations is to combine two gratings to have another one.
When the first grating is described with the function $\Phi^{(1)}$ and the second $\Phi^{(2)}$, the new operation is described using the function 
$\Phi_c=\Phi^{(1)}+\Phi^{(2)}$, with simple change of the image sent to the SLM. This combination does not change the obtained results trivially.
However, if one of the combining gratings is a constant phase it is possible to see that the action
of this combination is to multiply all the column of the matrix ${\mathcal{M}}$ defined by this grating by this constant phase:
\begin{equation}
 C_{jl} \stackrel{\rm composition}{\longrightarrow}e^{i\theta}C_{jl}.
\label{compositionConst}
\end{equation}

In the case of a saw-tooth grating with $\varphi = 2\bar{m}\pi$ being combined to a previous grating (described by $\Phi^{(1)}(y')$), the effect is to translate the grating column by $\bar{m}$. Calling $C^c_{j}$ the coefficient of the composed grating $\Phi_c(y')$, this can be shown as follows:
\begin{align}
C_j^c&=\int_{-\frac{T}{2}}^{\frac{T}{2}} e^{i\Phi_c(y')}\,e^{\frac{-2\pi ij}{T}\,y'}\,dy'\nonumber \\
&=\int_{-\frac{T}{2}}^{\frac{T}{2}} e^{\frac{2\bar{m}\pi}{T}y'}\,e^{i\Phi^{(1)}(y')}\,e^{\frac{-2\pi ij}{T}\,y'}\,dy'\nonumber\\
&=\int_{-\frac{T}{2}}^{\frac{T}{2}} e^{i\Phi^{(1)}(y')}\,e^{-i\frac{2\pi}{T}(j\,-\,\bar{m})\,y'}\,dy'\nonumber\\
&=C_{(j\,-\,\bar{m})},
\label{compositionSawTooth}
\end{align}
where $C_{j}$ is the coefficient of the $\Phi^{(1)}(y')$ grating.

These two possibilities can contribute to achieve the wanted operation without the necessity to look for different phase gratings. However, they are limited
by the maximum modulation of the SLM, which means that the maximum value of $\varphi$ of the combined gratings cannot exceed this maximum modulation, otherwise 
the implemented composed operations are not given by \eqref{compositionConst} or \eqref{compositionSawTooth}.

\subsubsection{More general maps}

We have considered transformations such as ${\rho} \rightarrow {\mathcal{M}}{\rho}{\mathcal{M}}^{\dagger}$,
now we address the problem of operations given by maps of the more general form ${\rho} \rightarrow \sum_i \bar{K}_i{\rho} \bar{K}_i^{\dagger}$.
Using the automated characteristic of the SLM this can be made easy if the source of the initial state is not deterministic in time,
similar as described in details in \cite{BRENOSLM}. In this reference, photonic qudits are prepared in the path variables and the SLM' screen is divided in $d$ regions aligned with the paths. In each of those regions, it is used a constant phase function.
By changing the phase masks in the SLM screen, with the \textit{i}-th mask being in display by a fraction $t_i$ of the detection time, one can see that $t_i$ is proportional
to $p_i$, the probability of implementing the operation  ${\mathcal{M}}_i$, represented by the \textit{i}-th phase mask. It means that in this case
the transformation in the initial state is:
\begin{align}
{\rho} \rightarrow &\sum_i p_i {\mathcal{M}}_i {\rho} {\mathcal{M}}_i^{\dagger} =\sum_i \bar{K}_i{\rho} \bar{K}_i^{\dagger},
\label{MapasGerais}
\end{align}
where $\bar{K}_i = \sqrt{p_i}{\mathcal{M}}_i$. Comparing \eqref{MapasGerais} with \eqref{Kraus} in sec. \ref{sec:Basic}, we can see that this proposal can be used to simulate open quantum systems dynamics, if the $\bar{K}_i$ represent correctly the kraus operators of such dynamics. In reference \cite{BRENOSLM} this was done with slit state qudits and the diagonal-only transformation in such encoding made that some maps whose Kraus decomposition have non-null off-diagonal elements could not be simulated. With this proposal and multipath encoded qudit state this limit is overcome; then, other open quantum systems dynamics simulations can be done if one find the correct phase gratings to implement the Kraus operatros of such dynamics.

Another way of having the convex sum of operators acting on the initial state, as in \eqref{MapasGerais}, is to use a random number generator with the correct distribution to define the instants in which the operations implemented by the SLM are changed. 

\subsection{Effects of pixelation}

\begin{figure}[h!]
 \includegraphics[width = 9.2cm, height=3.5cm]{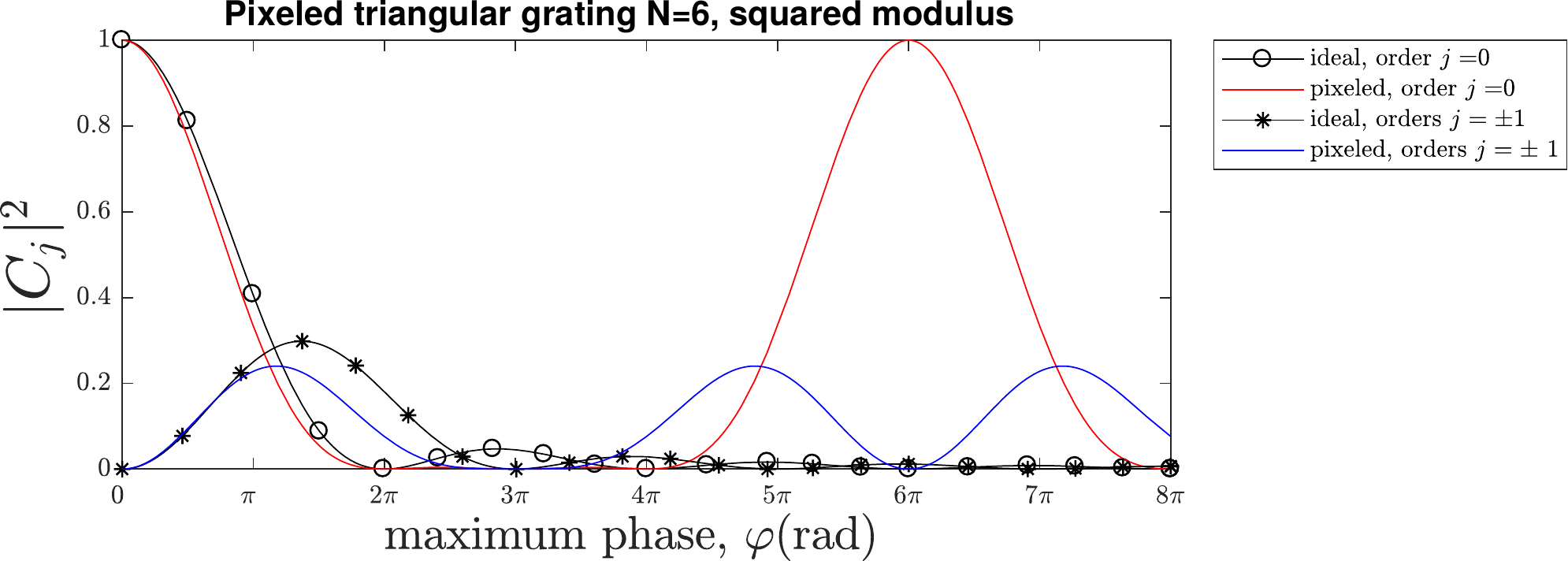}
\caption{Comparison of $|C_j|^2$ between the pixeled triangular grating and ideal triangular grating for the orders $j=\pm1$ and $j=0$ with $N=6$. It can be seen that these coefficients modulus present similar values that the idealized ones when $\varphi \approx 0$. However, near $\varphi = 6\pi$ the orders have a very different behaviour than that of the idealized ones with $C_j = \delta_{j0}$, where $\delta_{lj}$ is the Kronecker delta function.This anomalous behaviour have period $N\pi$.}
\label{Triangpix_mod6}
\end{figure}

\begin{figure}[h!]
 \includegraphics[width = 9.2cm, height=3.4cm]{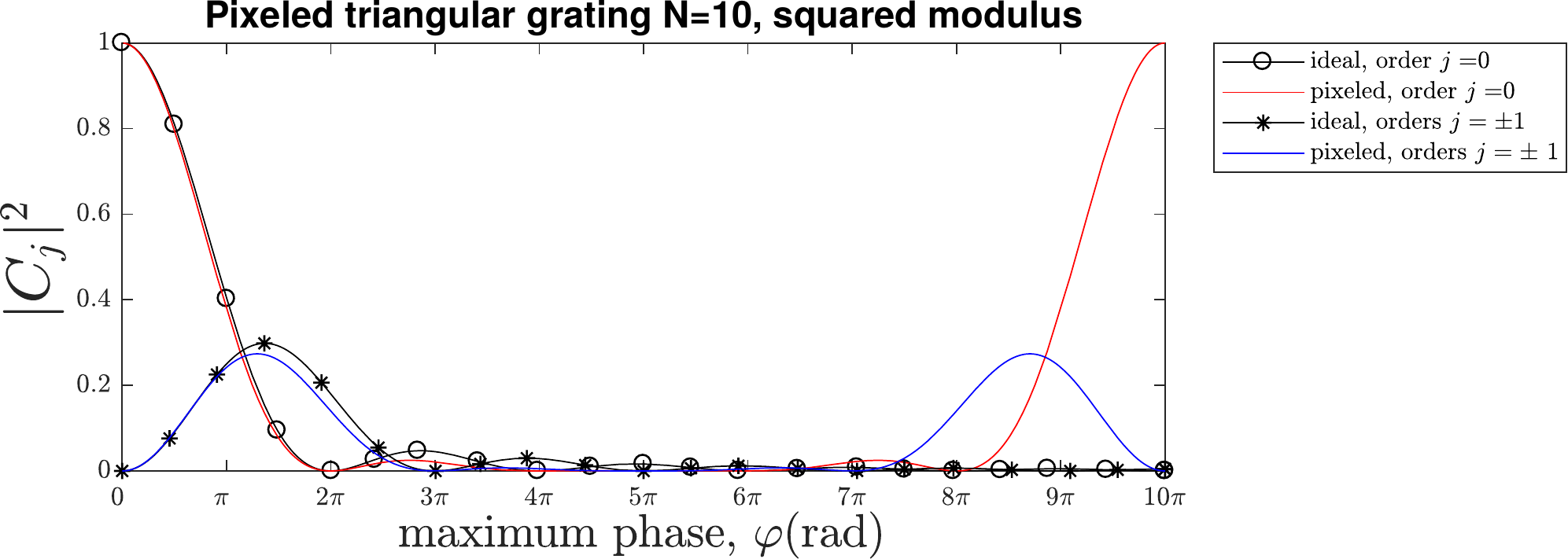}
\caption{Comparison of $|C_j|^2$ between the pixeled triangular grating and ideal triangular grating for the orders $j=\pm1$ and $j=0$ with $N=10$. The plot goes from $\varphi=0$ to
$\varphi=10\pi$ in order to make it possible to see the anomalous behaviour near $\varphi = 10\pi$, where $C_j = \delta_{j0}$, $\delta_{lj}$ being the Kronecker delta function. For the region $\varphi \approx 0$, the coefficients have similar modules than the one of the idealized grating.}
\label{Triangpix_mod10}
\end{figure}

\begin{figure}[h!]
 \includegraphics[width = 9.2cm, height=3.4cm]{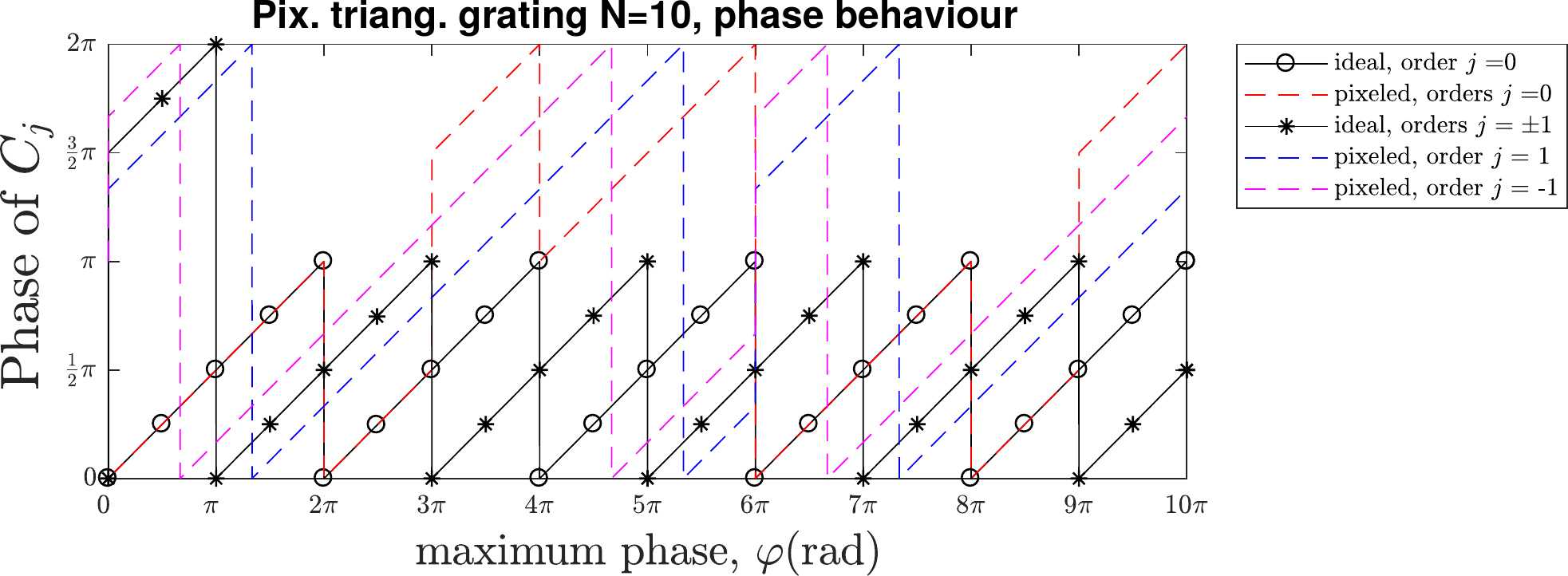}
\caption{Phase behaviour for the coefficients of the $N=6$ pixeled triangular function, compared with the idealized triangular function's coefficients for orders $j=0$ and $j=\pm1$. The linear dependence on $\varphi$ is mantained, but it is possible to see a difference due to the pixelization, making $C_j\neq C_{-j}$. The graphic is plotted to $\varphi = 10\pi$ as the $|C_j|^2$ plot, in Fig \ref{Triangpix_mod6}. For $\varphi \approx 3\pi$, it is possible to see that, for almost all values of $\varphi$ in the range $\{3\pi,8\pi\}$ there is a difference of approximately $\pi$ between the two cases.} 
\label{Triangpix_phase6}
\end{figure}

\begin{figure}[h!]
 \includegraphics[width = 9.2cm, height=3.4cm]{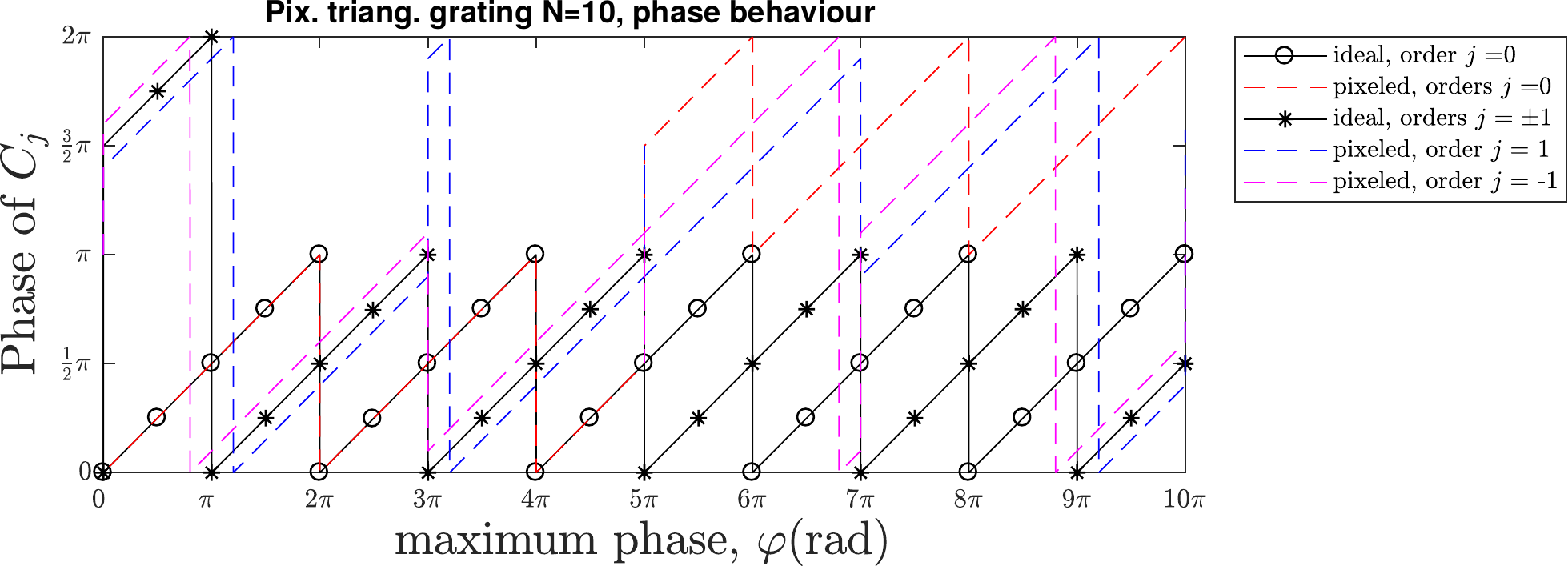}
\caption{Relative phase between the $N=10$ pixeled triangular function's coefficient and the path state component of the initial state. This plot compares the pixeled function results with the idealized triangular function's coefficients for orders $j=0$ and $j=\pm1$. The linear dependence on $\varphi$ is maintained, but it is possible to see a difference due to the pixelization that makes $C_j\neq C_{-j}$. The graphic is plotted to $\varphi = 10\pi$ as the $|C_j^2$ plot, in Fig \ref{Triangpix_mod10}. For $\varphi \approx 5\pi$, it is possible to see that, for almost all values of $\varphi$ in the range $\{5\pi,10\pi\}$ there is a difference of approximately $\pi$ between the two cases.}
\label{Triangpix_phase10}
\end{figure}

The phases applied by the SLM are controlled by a discrete grayscale or a discrete voltage scale, what can impinge differences. The differences caused by these discretizations will depend on the number of values that the grayscale or voltage scale can assume and
the value of maximum modulation ($\varphi$) used in the gratings. We will not consider it here. However, it might be important in some applications to
analyse carefully the problems that might occur because of that phase scale discretization. Another source of discretization imposed by the SLM is due to the fact that its secreen is pixeled. This means that the SLM cannot
apply continuous phase function in the plane $\mathbf{x}$: for each pixel, the applied phase is constant.
In reality, the phase is not applied by the whole pixel, but by a fraction of its area, defined by the SLM fill factor. Since it is common to have fill factors $f>0.9$,
and even $f\approx1$, we will not consider this effect.  The pixelation, however, imposes a discretization in the phase function $\Phi(y')$
 that cannot be neglected for some masks.
To see the effect of the pixelation, lets consider a triangular phase grating, in the form  given by \eqref{Triangular}.
The actual implementation in the SLM is given by:
\begin{align}
\Phi(y')= \begin{cases}
e^{i\frac{2n\varphi}{N}}, & 0 \leq n \leq \frac{N}{2}-1; \\
e^{\left(i\frac{2n\varphi}{N}-2\varphi\right)}, & \frac{N}{2} \leq n \leq N-1,
\end{cases}
\label{Triangpix}
\end{align}
where $n$ defines each pixel, labelled from $n=0$ to $n=N-1$ for a period of $N$ pixels. The function on \eqref{Triangpix} gives the following coefficients:
\begin{widetext}
\begin{align}
C_0 &= \frac{2}{N}\sinc{\left(\frac{j}{N}\right)}e^{i\frac{\varphi}{2}}\frac{\sin{\left(\frac{\varphi}{2}\right)}}{\sin{\left(\frac{\varphi}{N}\right)}}\cos{\left(\frac{\varphi}{N}\right)}, \\[3mm]
C_j&=\frac{\sinc{\left(\frac{j}{N}\right)}}{N}e^{i\frac{\varphi-\pi j}{2}}\left[e^{-i\frac{\varphi}{N}}\frac{\sin{\left(\frac{\varphi-\pi j}{2}\right)}}{\sin{\left(\frac{\varphi-\pi j}{N}\right)}} + e^{i(\frac{\varphi}{N}-\pi j)}\frac{\sin{\left(\frac{\varphi+\pi j}{2}\right)}}{\sin{\left(\frac{\varphi + \pi j}{N}\right)}}\right].
\label{CTriangPix}
\end{align}
\end{widetext}
These results are different than the ones in \eqref{CTriangular}, what imposes a modification in the idealized results that can be severe. One can note that the $C_0$ coefficient is $1$ for some values of $\varphi$ that depends on $N$ (see Figs. \ref{Triangpix_mod6} and \ref{Triangpix_mod10}). It is interesting to note that this severe modification happens for higher values of $\phi$, as $N$ increases. Another important modification between these results and the idealized ones is that, for this grating, $C_j$ is no longer identical to $C_{-j}$. This can be seen in the Figs. \ref{Triangpix_mod6} to \ref{Triangpix_phase10}, that summarize these effects for $N=6$ and $N=10$.
This means that the pixelation modifies the relative phase of the coefficients and their modulus, changing the actual implementation usually decreasing
the diffraction efficiency, for some values of $\varphi$ -- that goes to $\infty$ as $N$ increases.  Actually, the expression in
\eqref{CTriangPix} tends to the expressions of the idealized function [\eqref{CTriangular}] if $N \rightarrow \infty$. This limit has to be carefully considered:
the separation between the orders of diffraction is proportional to $\frac{1}{T} = \frac{1}{Nl}$, where $l$ is the size of the pixel in the direction of the 
periodic behaviour of the grating. This means that only increasing the number of pixels is not the best way to have a good experimental result, since the 
separation between the orders is important to have the qudit well encoded. It is important to have $N\rightarrow \infty$ while $Nl \rightarrow T = \text{constant}$, 
which means smaller pixels. This implies that in order to have a better agreement of the pixeled and idealized functions, it is better to have a SLM with smaller pixels. 

It can be seeing in Fig \ref{Triangpix_mod6} to \ref{Triangpix_phase10} that, for intervals of $\varphi$ in which $\frac{\varphi}{N} \in \{0,\pi/6\}$ for the cases considered, the difference of the pixeled and ideal grating for the modulus of the coefficients can be negligible, while the phase of $C_j$ and $C_{-j}$ are now different. 

This result can be generalized to other linear functions of $y$, such as the saw-tooth grating. In this case, for example, a $\varphi = 2\pi$
grating implies a $C_1 = 0.96$, for $N = 12$, instead of $C_1=1$, for the idealized case. This means that for the simpler linear gratings 
this effect can be made of order of less than $5$ per cent. In our example, we found an analytical form to describe the pixelation effects. However, depending
 on the mask this can be hard; the coefficients and their deviation to the idealized case may have to be numerically calculated. 

\section{\label{sec:Conclusion}Conclusion and perspectives}

In this work, we present a proposal to have
more general automated operations on multipath photonic pure qudits encoded in one direction. This is done by using periodic phase gratings on a phase-only SLM and a interferometer that merges the initial path state components into one alone, while the transformed state will be now  encoded in the orthogonal direction. It is important that the state is prepared with an amplitude that is eigenfunction of the Fourier Transform. Since the operation is controlled by the function used in the programmable SLM, this proposal is completely automated. This proposal overcomes the limitation
for implementing non-diagonal operations present in the slit states while preserving the final state and the dimension limitation present in polarization encoded qudits. In addition, since there is a wide range of different phase gratings to be used, there are several transformations possible with this proposal. This method can be used to make sequential
operations, creating new possibilities for experimental studies in fundamental quantum theory and quantum computation protocols.
We also considered the effects of pixelation due to the SLM screen, showing how it affects the results for some phase gratings.

This work can be complemented by the use of photonic chips or optical fibres to substitute the interferometer for a better control and by a study of the generality of this approach in order to know what operations can be made with this proposal. It would be important to search for an algorithm to find the correct phase function for a given desired matrix operation.

Important operations, such as permutations and projections, however, can already be done with this proposal, in an automated way,
using the gratings showed in this paper, bypassing important difficulties already found in discrete variable quantum optics experiments.

\acknowledgments
The authors thank J. Cond\'e for the discussions, comments and interest in this proposal and M. T. Cunha, R. Rabelo and A. Cabello for useful discussions and important remarks. This research was supported by CAPES, CNPq and FAPEMIG.

\bibliographystyle{unsrtnat}
\bibliography{bibliography}

\end{document}